\RequirePackage{lineno}
\documentclass[a4paper,11pt,showkeys,nofootinbib]{revtex4}
\usepackage{graphicx,url}
\usepackage{soul,xcolor}
\usepackage{amsmath, amssymb, setspace, color}

\begin{document}
\title{Whole body counter surveys of Miharu-town school children for four consecutive years after the Fukushima NPP accident}
\author{Ryugo S. Hayano}
\thanks{Correspondence should be addressed: R. Hayano, (\url{hayano@phys.s.u-tokyo.ac.jp}).}
\affiliation{Department of Physics, The University of Tokyo, 7-3-1 Hongo, Bunkyo-ku, Tokyo 113-0033, Japan}
\author{Masaharu Tsubokura}
\affiliation{Division of Social Communication System for Advanced Clinical Research, Institute of Medical Science, The University of Tokyo, 4-6-1 Shirokanedai, Minato-ku, Tokyo 108-8639, Japan}
\author{Makoto Miyazaki}
\affiliation{Department of Radiation Health Management, Fukushima Medical University, Hikariga-oka, Fukushima 960-1295, Japan}
\author{Hideo Satou, Katsumi Sato, Shin Masaki, Yu Sakuma}
\affiliation{Hirata Central Hospital and Research institute of radiation safety for disaster recovery support, Hirata village, Fukushima 963-8202, Japan}

\begin{abstract}
Comprehensive whole-body counter surveys of Miharu town school children have been conducted for four consecutive years, in 2011--2014. This represents the only long-term sampling-bias-free study of its type conducted after the Fukushima Dai-ichi accident.  For the first time in 2014, a new device called the Babyscan, which has a low $^{134/137}$Cs MDA of $< 50$~Bq/body, was used to screen the children shorter than 130~cm. No child in this group was found to have detectable level of radiocesium. Using the MDAs, upper limits of daily intake of radiocesium were estimated for each child. For those screened with the Babyscan, the upper intake limits were found to be $\lesssim 1$~Bq/day for $^{137}$Cs.
Analysis of a questionnaire filled out by the children's parents regarding their food and water consumption shows that the majority of Miharu children regularly consume local and/or home-grown rice and vegetables. 
 This however does not increase the body burden.
\end{abstract}
\maketitle

\section{Introduction}
Nearly four years have elapsed since the start of the Fukushima Dai-ichi nuclear power plant accident~\cite{tanaka}, which began in March, 2011. Because of the large amounts of radionuclides deposited on soil and water in Fukushima Prefecture and surrounding regions of Japan, the risk of serious internal radiation exposure for residents was of great concern initially, but most data accumulated and disseminated so far have consistently shown that the internal contamination for the overwhelming majority of residents has fortunately been so low as to be undetectable~\cite{unscear2013}. The data include, for example, whole-body-counter surveys~\cite{hayano2013,nagataki,tsubo1,tsubo2}, duplicate-diet studies~\cite{coop},  the inspection of ``all rice in all rice bags'' harvested in Fukushima (2012-2014)~\cite{rice}.

\begin{figure}
\includegraphics[width=0.5\columnwidth]{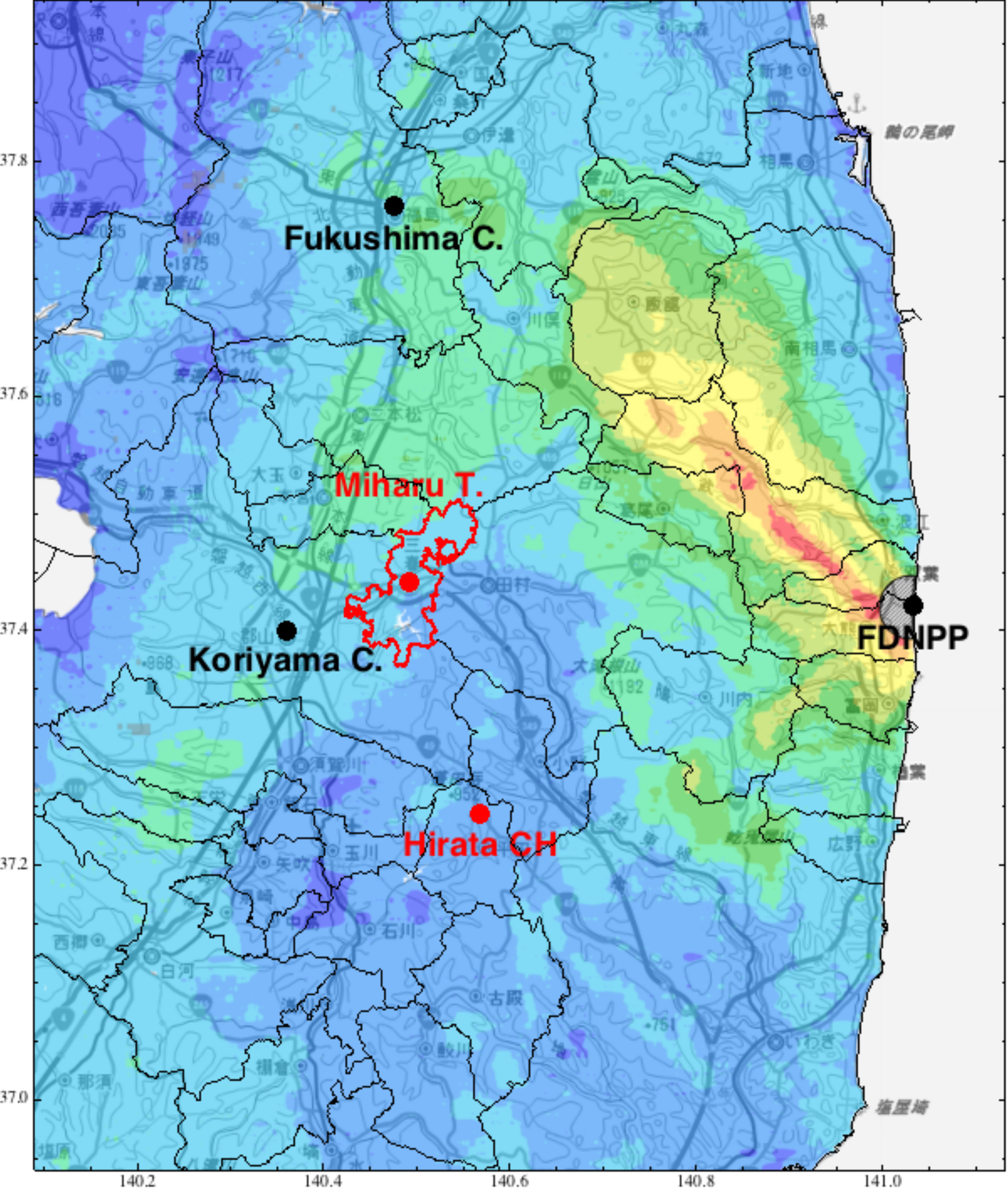} 
\includegraphics[width=0.3\columnwidth]{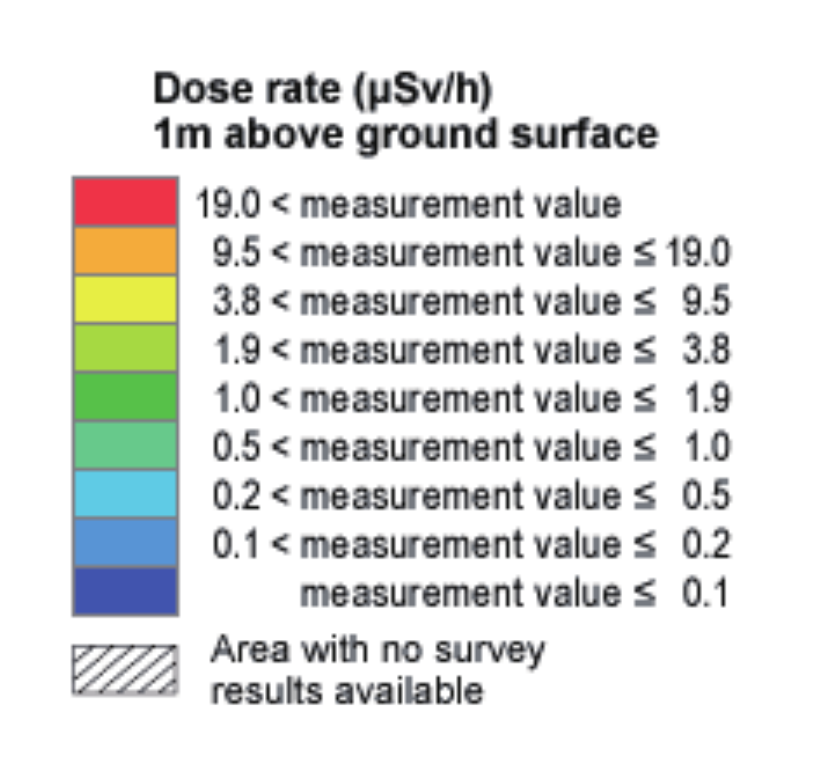}
\caption{\label{fig:map} A map of the air dose rate ($\mu$Sv/h) estimated from airborne monitoring, as of November 19, 2013~\cite{map}. The locations of the Fukushima Dai-ichi nuclear power plant (FDNPP), the town of Miharu, Fukushima City, Koriyama City, and Hirata central hospital (Hirata CH) are indicated.}
\end{figure}

This does not mean that the lives of Fukushima residents have returned to normal, however. A survey conducted by Fukushima City in May, 2013, \cite{fukushimacity} indicated that about 80\% of Fukushima City residents are still concerned about the risk of internal radiation exposures from food. Another study showed that about three quarters of the parents of Minamisoma school children buy only non-Fukushima produced foodstuffs at the grocery store, out of fear of radioactive contamination~\cite{minamisomaschool}.

In this regard, the results of the whole-body-counter surveys of Miharu-town school children, conducted for four consecutive years (2011--2014) with high ($>90\%$) coverage provide valuable sampling-bias-free information, are worth reporting, particularly because the majority of the children are regularly consuming local/home-grown foodstuffs, as will be shown later.

The Town of Miharu (population $\sim 18,000$), located about 50~km west of Fukushima Dai-ichi, is in the suburbs of Koriyama city (population $\sim 330,000$). As shown in Fig.~\ref{fig:map}, the soil contamination level, inferred from the airborne radiation monitoring results (November 2013) is moderate in the southern part of the town, and is slightly higher in the northern part.  The percentage of agricultural households is about 20\%. There are six municipally-operated primary schools, attended by approximately 800 children in all, and two secondary schools, with about 500 children in total.  In the fall of 2011, the Miharu-town school board decided to have all the school children tested for internal radiation exposures. This was continued in 2012, 2013 and 2014.

The results of the 2011-2013 internal-contamination surveys have been reported in Ref.~\cite{miharu}. Although the results of the questionnaires  of 2012-2013 indicated that approximately 60\% of the children had been regularly eating local or home-grown rice, in 2012 and 2013 no child was found to exceed the $^{137}$Cs detection limit of 300 Bq/body.

This present paper reports the results of the 2014 internal-contamination surveys, in which we used a newly-developed whole-body counter called Babyscan~\cite{babyscan}, having a detection limit of $< 50$~Bq/body, for children with a height of less than 130~cm. 

The study was endorsed by the Miharu-town school board, and was approved by the Ethics Committee of the University of Tokyo.

\begin{table*}
\caption{Breakdown of WBC measurement results for Miharu Town school children (age 6-15). The numbers of enrolled children were published by the school board of Miharu Town, and  some discrepancies with the numbers of children actual attending schools when the WBC measurements were carried out may exist. \\
i) Measured between Nov 24, 2011 and Feb 2, 2012. \underline{No change of clothes}\\
ii) Measured between Sep 3, 2012 and Nov 8, 2012. \\
iii) Measured between Sep 2, 2013 and Nov 29, 2013.\\
iv) Measured between March 24, 2014 and Nov 28, 2014.\\
a) Based on the Miharu school board statistics as of August 25, 2011.\\
b) Based on the Miharu school board statistics as of April 1, 2012.\\
c) Based on the Miharu school board statistics as of April 1, 2013.\\
d) Based on the Miharu school board statistics as of April 1, 2014.}
\label{tab:miharu}
\begin{tabular}{l|r r r r r}
\hline
&Enrolled & \multicolumn{1}{l}{Measured} &  Coverage&Radiocesium &\multicolumn{1}{l}{Detection}\\
& &  (with Babyscan)&& \multicolumn{1}{l}{detected}& \multicolumn{1}{l}{percentage}\\
\hline
2011$^{i)}$& 1,585$^{a)}$& 1,494~~~(0)& 94.3\%&54\footnote{As discussed in Ref.~\cite{hayano2013}, some of these detections may have been caused by surface (clothes) contamination.} & 3.6\%\\
2012$^{ii)}$& 1,413$^{b)}$& 1,383~~~(0)& 97.9\%&0&0.0\%\\
2013$^{iii)}$& 1,381$^{c)}$ & 1,338~~~(0)&96.9\% & 0&0.0\%\\
2014$^{iv)}$& 1,315$^{d)}$ & 1,265(360) & 96.2\% & 0&0.0\%\\
\hline
\end{tabular}
\end{table*}

\section{Details of the 2014 surveys}
\begin{figure}
\begin{minipage}{0.48\textwidth}
\includegraphics[width=\columnwidth]{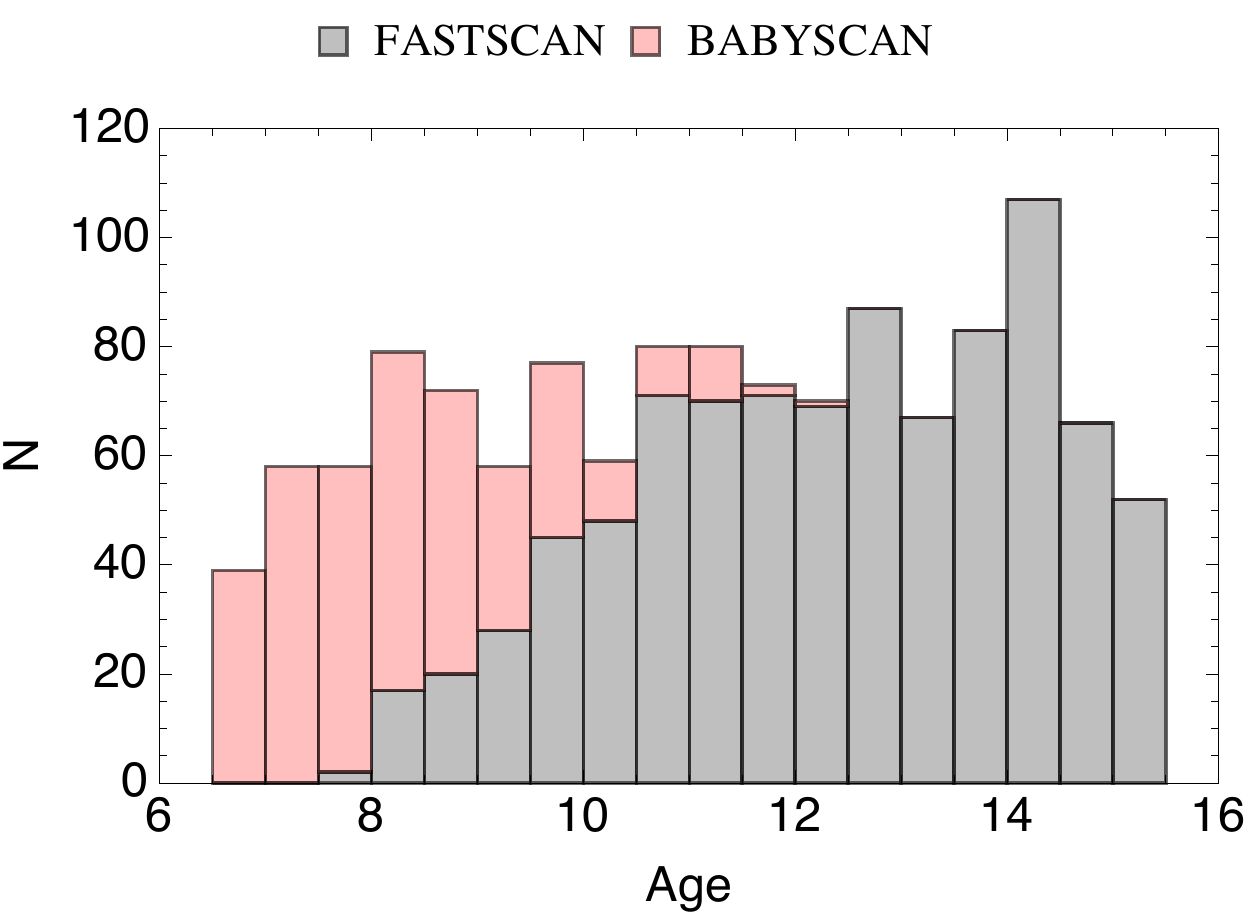}
\caption{\label{fig:age} The age distribution of the subjects. The gray (pink) bars are for those measured with the Fastscan (Babyscan).}
\end{minipage}
\hfill
\begin{minipage}{0.48\textwidth}
\includegraphics[width=\columnwidth]{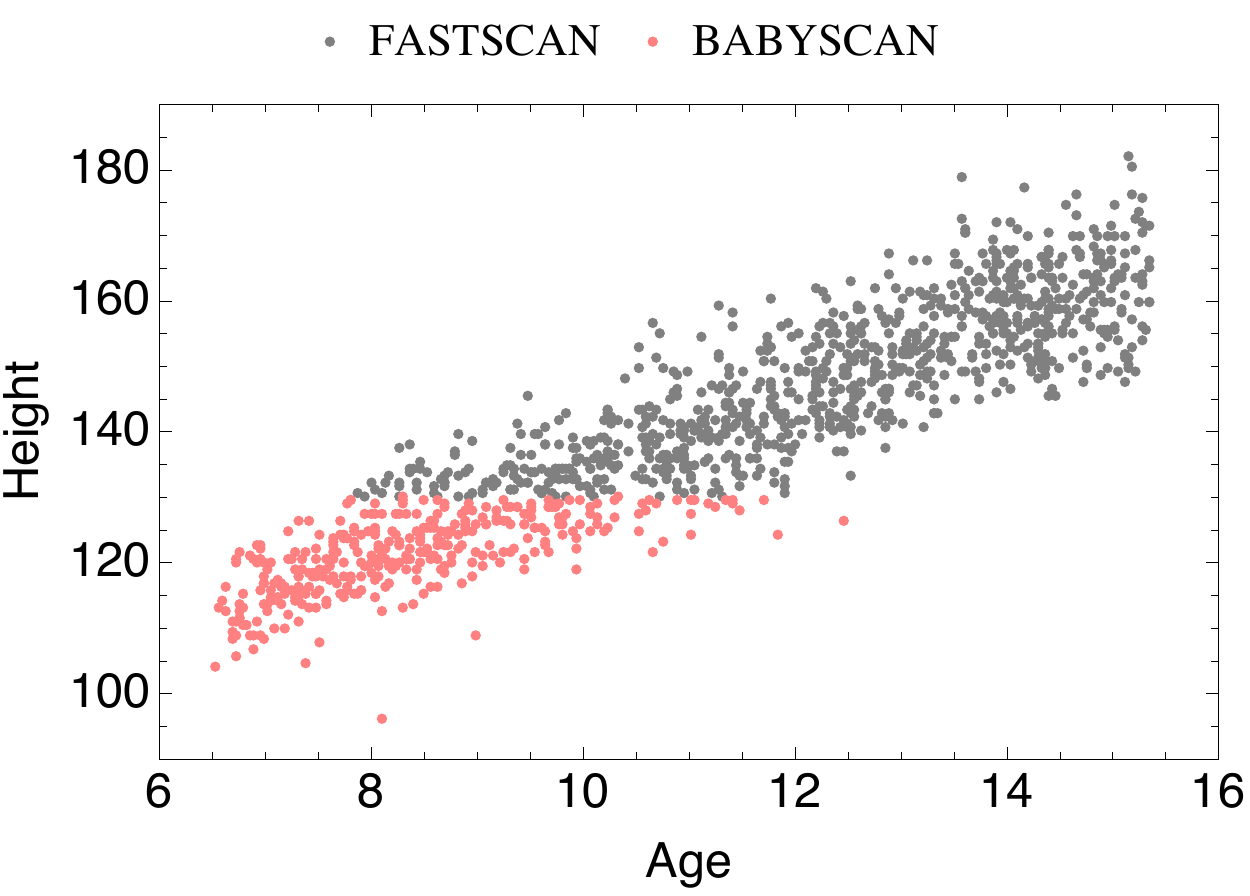}
\caption{\label{fig:age_vs_height} The distribution of age vs height of the subjects. The subjects shorter than 130~cm were scanned with the Babyscan.}
\end{minipage}
\end{figure}

In 2014, 1,265 children from Miharu Town between the ages of 6 and 15 (Fig.~\ref{fig:age}) were scanned for of $^{134}$Cs and $^{137}$Cs using whole-body counters at the Hirata central hospital (Fig.~\ref{fig:map}).  In this group, children taller than 130~cm were measured with a standing-type whole-body counter (Fastscan Model 55	2251, Canberra Inc.). The nominal detection limits were 300 Bq/body for both $^{134}$Cs and $^{137}$Cs following a 2-minute scan. Those shorter than 130~cm were measured with a newly-developed whole-body counter for small children (Babyscan, Canberra Inc.), having nominal detection limits of 50 Bq/body following a 4-minute scan (Fig.~\ref{fig:age_vs_height}). In addition, before the WBC measurement, all of the parents of the participating children were asked to complete a questionnaire regarding their family's food and water consumption, the results of which will be discussed later.


\begin{figure}
\includegraphics[width=.5\columnwidth]{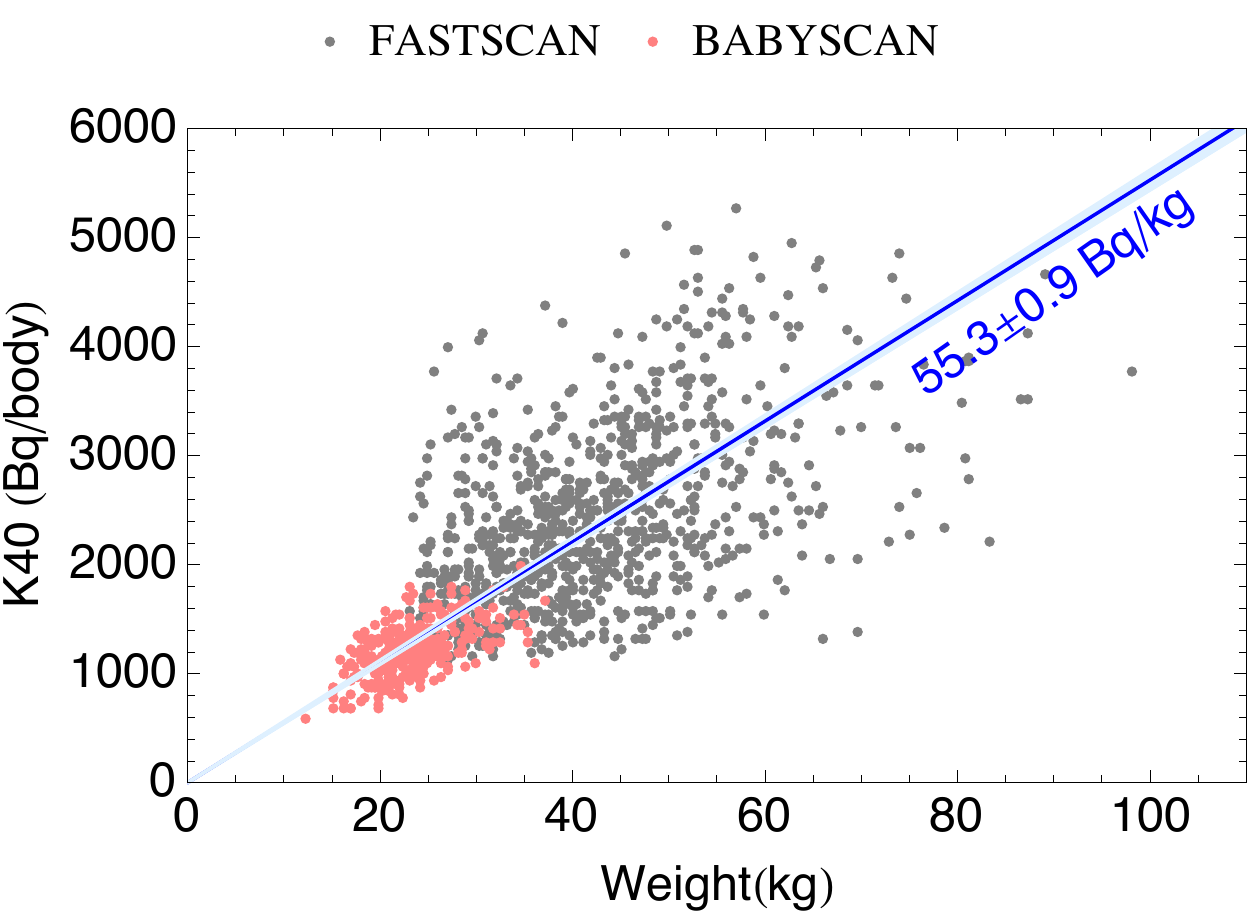}
\caption{\label{fig:weight_vs_k40} Weight (kg) vs  measured amount of $^{40}$K (Bq) in the subjects' body.}
\end{figure}

The quality of the data can be inferred from Fig.~\ref{fig:weight_vs_k40}, which shows the correlation between the subjects' weight and the measured activity of $^{40}$K in the body. Overall fit to the data points yielded a coefficient $55.3\pm 0.9$ Bq/kg, consistent with  the known amount of $^{40}$K in human body. The data obtained with the Babyscan show much smaller spread than those with the Fastscan, demonstrating the higher sensitivity (and hence lower detection limit) of the former. 

As in previous years, no child was found to exceed the nominal detection limits; this was true even for the small children measured with the Babyscan.  Table~\ref{tab:miharu} summarizes the 2014 results, together with those in 2011-2013.

\begin{figure}
\begin{minipage}{0.48\textwidth}
\includegraphics[width=\columnwidth]{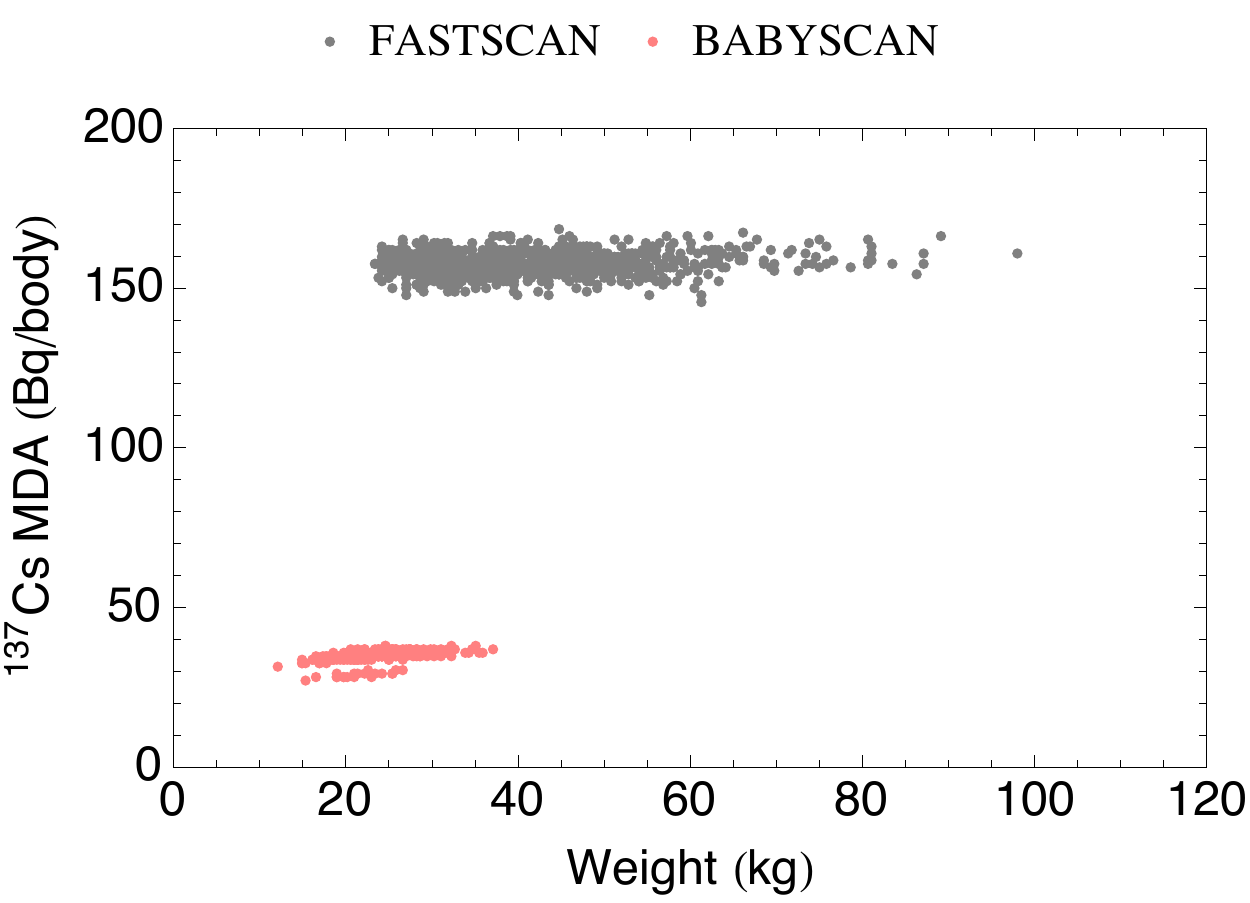}
\caption{\label{fig:mdabody} MDA body}
\end{minipage}
\hfill
\begin{minipage}{0.45\textwidth}
\includegraphics[width=\columnwidth]{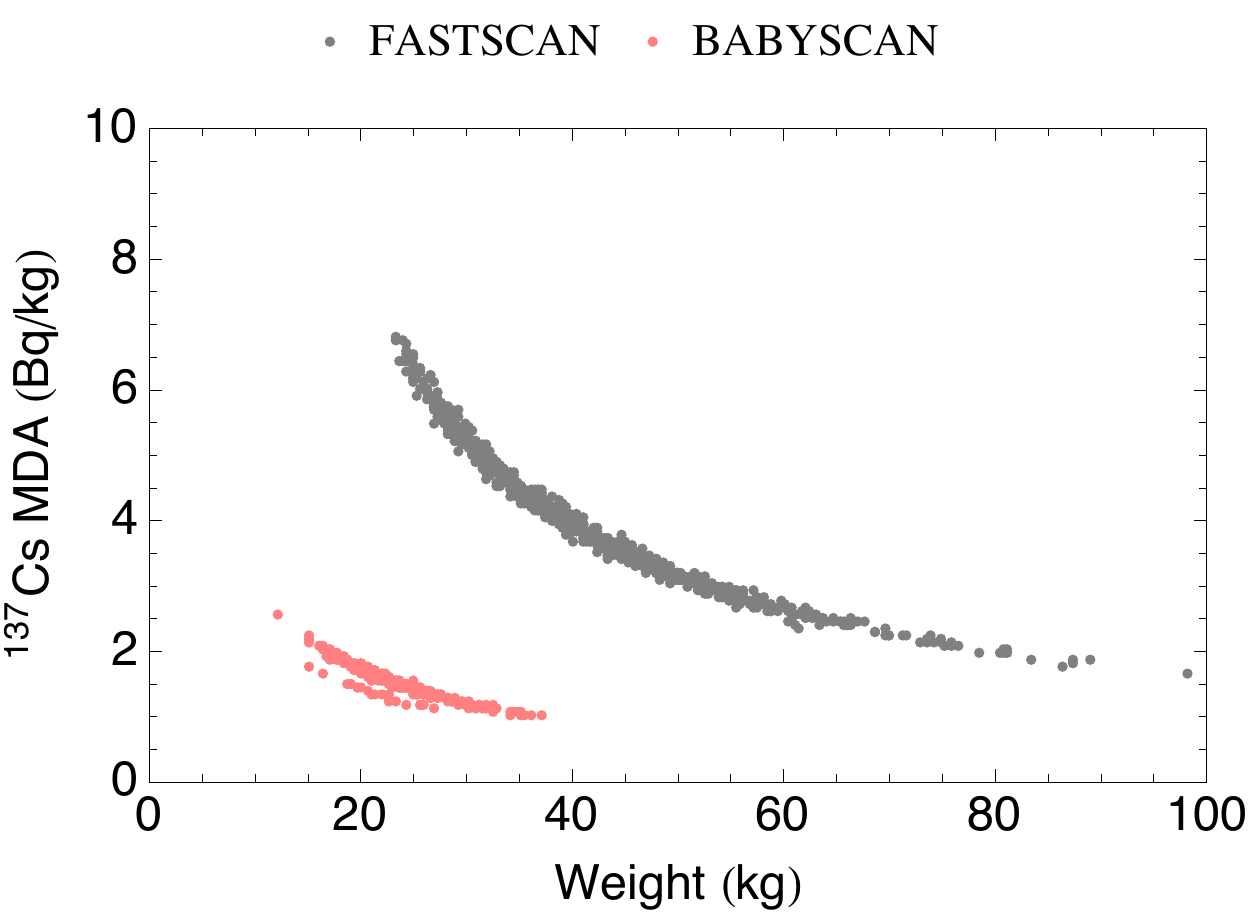}
\caption{\label{fig:mdaratio} MDA ratio}
\end{minipage}
\end{figure}

In more detail, Fig.~\ref{fig:mdabody} shows the minimum detectable activity~\cite{currie} (MDA for $^{137}$Cs (Bq/body) versus body weight, calculated for each subject from the statistical fluctuation of the gamma-ray spectrum at around 662 keV (the energy of the gamma-ray emitted in the $^{137}$Cs decay). As shown, the measured MDAs are lower than the nominal ones being used at the Hirata Central Hospital (300~Bq/body for Fastscan, and 50~Bq/body for Babyscan).
Fig.~\ref{fig:mdaratio} shows that the  $^{134}$Cs-to-$^{137}$Cs MDA ratios were about 0.9. For simplicity, we use the $^{137}$Cs MDAs in the following discussions.

\section{Discussion}
\subsection{Upper limit of daily ingestion of $^{137}$Cs}

\begin{figure}
\begin{minipage}{0.55\textwidth}
\includegraphics[width=\columnwidth]{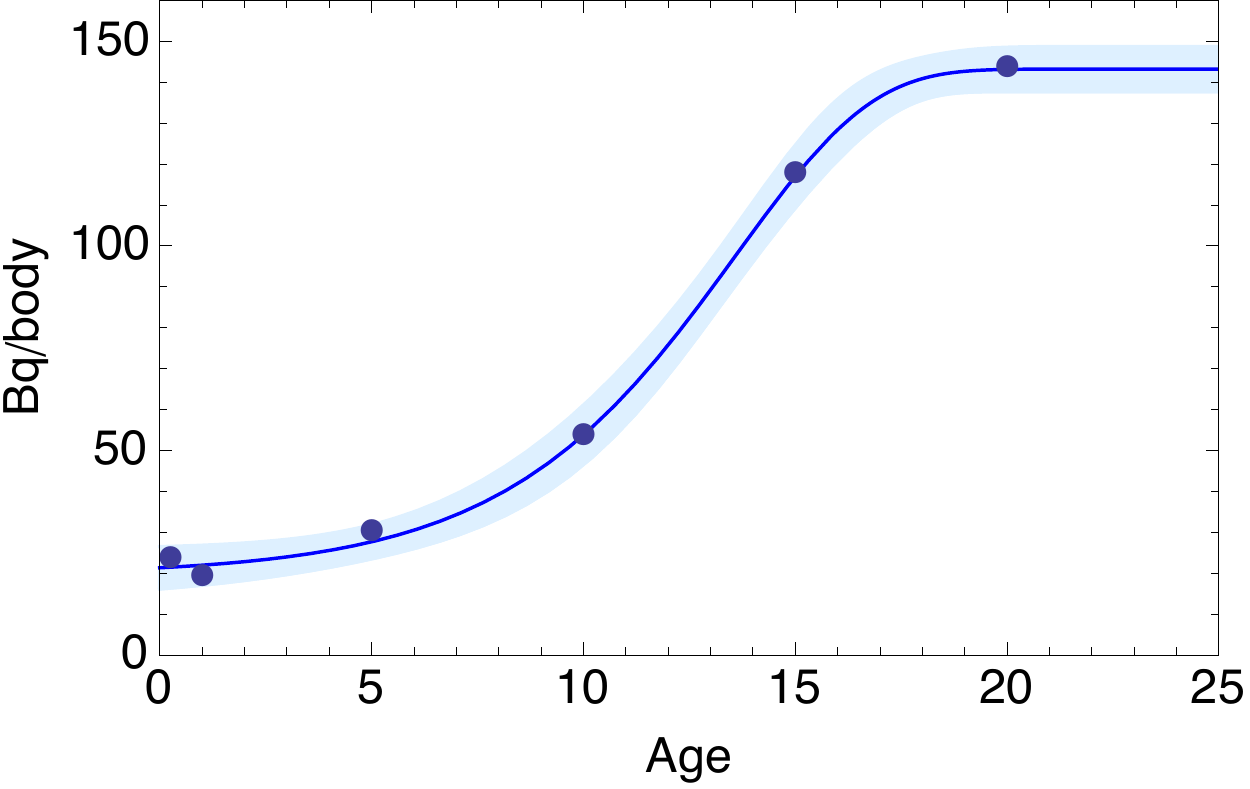}
\caption{\label{fig:mondal3} Circles: The equilibrium plateau values of $^{137}$Cs body burden calculated using Mondal3 software~\cite{mondal3}. Curve: interpolating curve, shown with a 1-$\sigma$ uncertainty band, obtained by fitting a cumulative Gompertz distribution to the Mondal3 outputs.}
\end{minipage}
\hfill
\begin{minipage}{0.4\textwidth}
\includegraphics[width=\columnwidth]{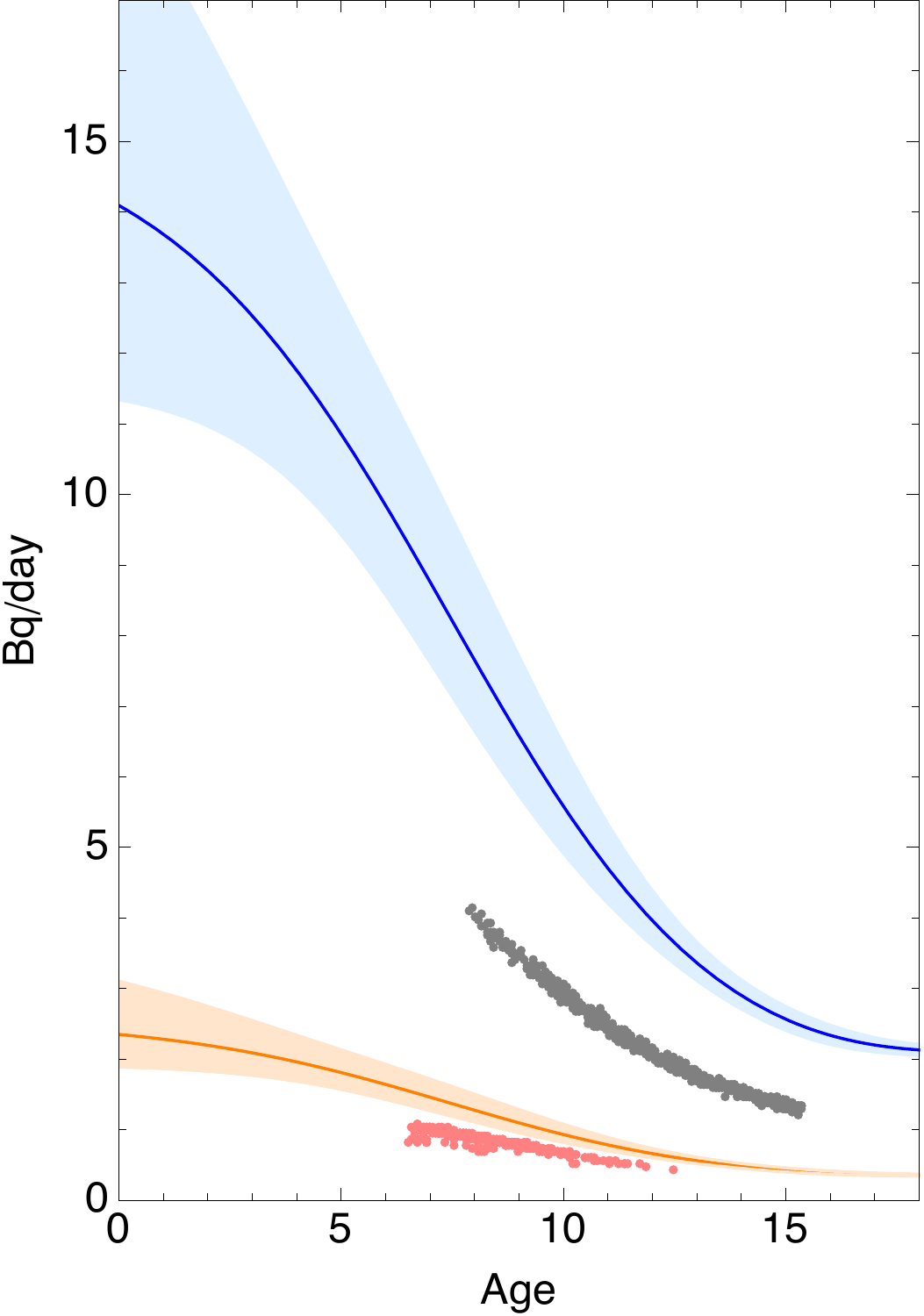}
\caption{\label{fig:intakeupperlimit} Blue (orange) curve: The age dependence of the estimated upper limit of $^{137}$Cs daily intake, for the detection limit of 300 (50) Bq/body. Gray (pink) dots: upper limits of $^{137}$Cs daily intake estimated using the actual MDA of the Miharu school children.}
\end{minipage}
\end{figure}

From the MDAs of the $^{137}$Cs, we estimate the upper limit of daily ingestion of $^{137}$Cs using a constant, long-term intake scenario. Assuming daily intake of 1~Bq, the equilibrium plateau values of $^{137}$Cs body burden were calculated using Mondal3 software~\cite{mondal3} for ages 3 months, 5 years, 10 years, 15 years and adults, and the calculated results were fitted with a smooth interpolating curve,  as shown in Fig.~\ref{fig:mondal3}\footnote{The function we used is the cumulative Gompertz distribution (the choice of this function being purely phenomenological). The fit yielded 
$21.3+121.9\left(
 1-e^{0.014 \left(1-e^{0.31 x}\right)}\right) $, which is shown in FIg.~\ref{fig:mondal3} together with a 1-$\sigma$ uncertainty band.}. Due to the shorter biological halflife, the plateau body burden is lower for younger children, and this is why we used the Babyscan to screen small children in this study.

Using the age-vs-body burden curve shown in Fig.~\ref{fig:mondal3}, we calculated the upper limit of $^{137}$Cs daily intake for the nominal detection limit of 300 (50) Bq/body,  shown in Fig.~\ref{fig:intakeupperlimit} in blue (orange) curves. Also shown in Fig.~\ref{fig:intakeupperlimit} are the estimated intake upper limits calculated from the actual MDAs of the Miharu school children. For those scanned with Babyscan, the estimated ingestion upper limit is 1 Bq/day (for 6-year old). This results in a committed effective dose below $\sim 4 \mu$Sv/year ($\sim  8\mu$Sv/year if the $^{134}$Cs contribution is included).  For this reason we conclude that the concomitant health risk is negligibly small.

\subsection{Analysis of the questionnaire}

In the questionnaire filled out by parents regarding their family's food and water consumption, we asked
\begin{itemize}
\item drinking water: choice of: \\
1. well water, 2. tap water, 3. bottled water
\item rice: choice of: \\
1. avoid Fukushima rice, 2. do not care about the source, 
3. buy Fukushima rice, 4. eat home/local rice

\item vegetables: choice of: \\
1. avoid Fukushima produce
2. do not care about the source, 
3. buy Fukushima produce, 
4. eat  home/local produce after being tested for radioactivity,
5. eat untested home/local produce
\end{itemize}

\begin{figure*}
\includegraphics[width=0.75\textwidth]{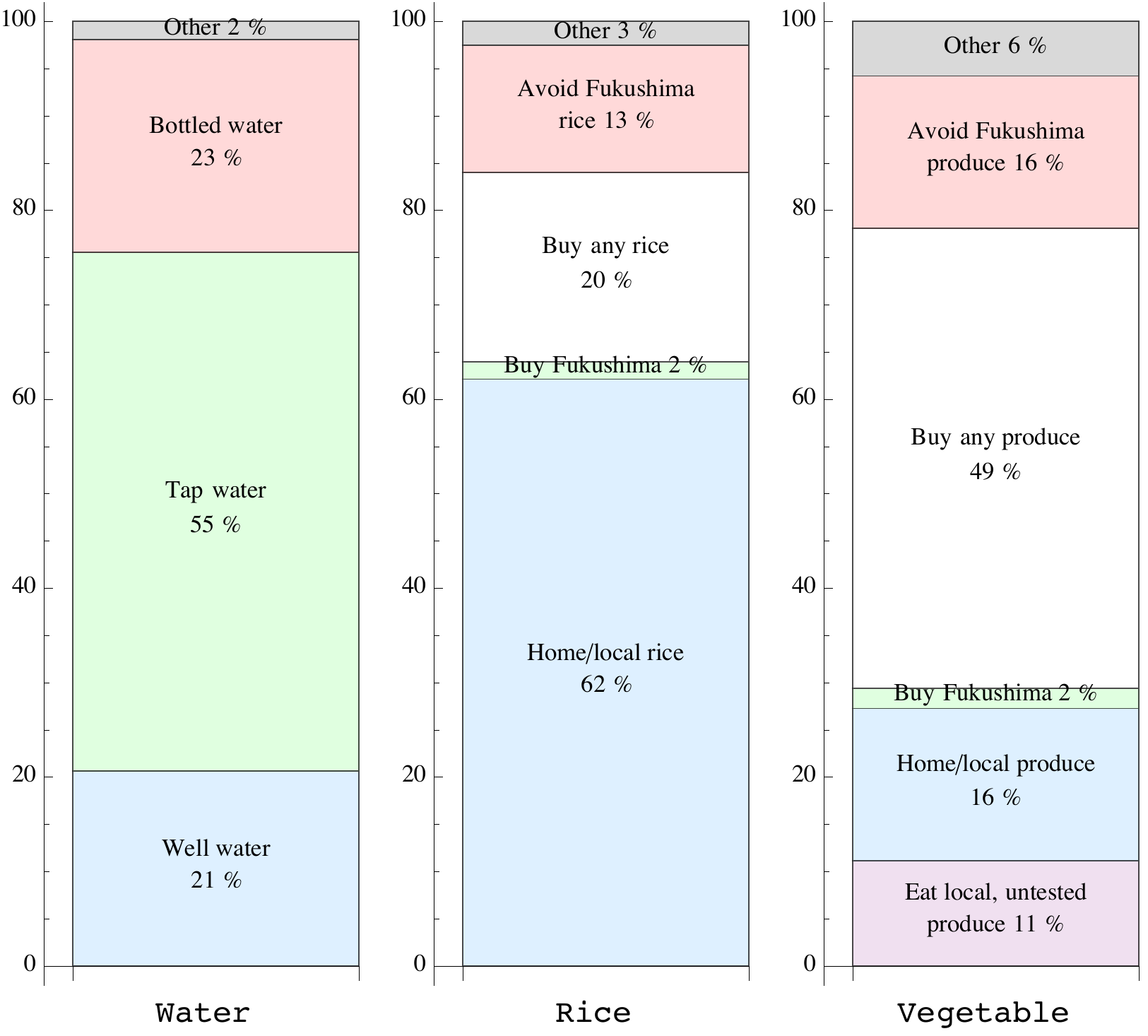}
\caption{\label{fig:waterricevegetable} Graph showing results of questionnaire concerning water, rice, and vegetable consumption of Miharu families.}
\end{figure*}

Fig.~\ref{fig:waterricevegetable} shows a summary graph, and Table \ref{tab:cross}  gives the actual numbers in the form of a cross table.

\begin{table*}
\caption{\label{tab:cross} The cross table summarising the questionnaire results (n=1146). From this table, 119 sheets which had multiple choices or no entry were excluded.}
\begin{footnotesize}
\begin{tabular}{l|rrrr|rrrrr}
\hline
&\multicolumn{4}{c}{Rice}&\multicolumn{5}{|c}{Vegetables}\\
&\multicolumn{3}{c}{$\leftarrow~~~~$Supermarket$~~~~\rightarrow$}& Home/Local&\multicolumn{3}{c}{$\leftarrow~~~~$Supermarket$~~~~\rightarrow$}&\multicolumn{2}{c}{Home/Local}\\
&\multicolumn{1}{l}{Avoid} &\multicolumn{1}{l}{Buy} &\multicolumn{1}{l}{Buy}& Tested&\multicolumn{1}{l}{Avoid} &\multicolumn{1}{l}{Buy}&\multicolumn{1}{l}{Buy}&Tested&Untested\\
&Fukushima & \multicolumn{1}{l}{any~~~~~~~~}& \multicolumn{1}{l}{Fukushima}&& Fukushima& \multicolumn{1}{l}{any~~~~~~~~} & \multicolumn{1}{l}{Fukushima}& &\\
\hline
 \text{Well water} & 13 & 25 & 4 & 195 & 27 & 91 & 7 & 61 & 51 \\
 \text{Tap water} & 84 & 154 & 14 & 387 & 105 & 349 & 14 & 104 & 67 \\
 \text{Bottled water} & 64 & 66 & 4 & 136 & 64 & 146 & 6 & 34 & 20 \\
\hline
\end{tabular}
\end{footnotesize}
\end{table*}

As shown, about 75~\% of Miharu families drink tap and well water, 80~\% eat Fukushima rice, and 75~\% eat Fukushima vegetables (some 10\% regularly eat home-grown or local vegetables that have not been tested for radioactivity).
 This is in sharp contrast to the situation in the city of Minamisoma, where about 3/4 of families answered that they avoid local/Fukushima produce~\cite{minamisomaschool}.

The current study supports previous ones which have suggested that the risk of significant internal radiation exposures among the population of Fukushima remains small, even among people who consistently eat locally-grown produce.

\subsection{Correspondence analysis of the water-rice-vegetable cross table}
Figure \ref{fig:waterricevegetable} shows that about 20\% of the families buy bottled water, and about 15\% avoid Fukushima produce, which appears to suggest that the choice of bottled water and avoidance of Fukushima produce may be correlated.

\begin{figure}
\setlength{\fboxsep}{0pt}
\setlength{\fboxrule}{0.5pt}
\fbox{\includegraphics[width=0.4\columnwidth]{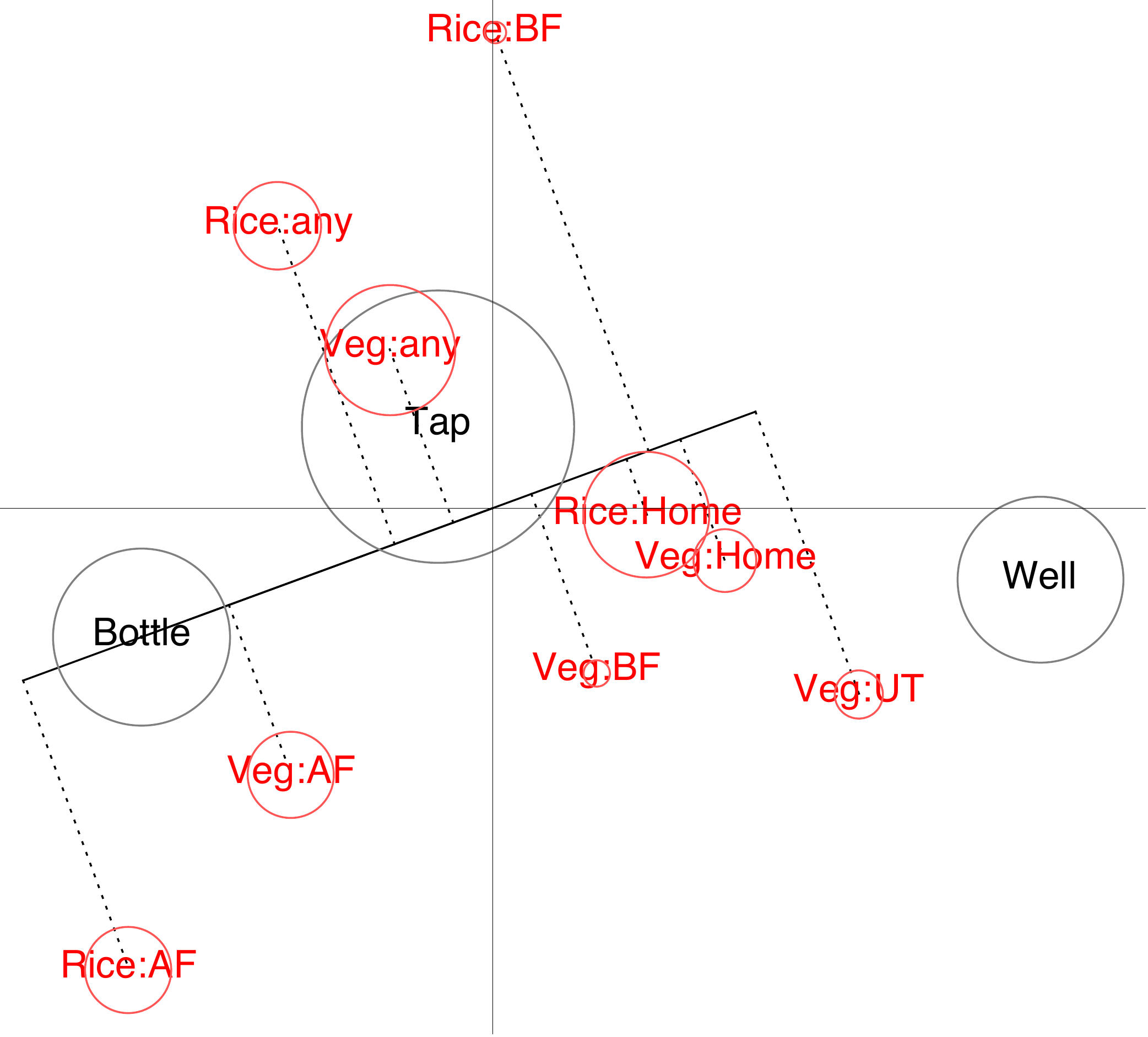}}
\fbox{\includegraphics[width=0.4\columnwidth]{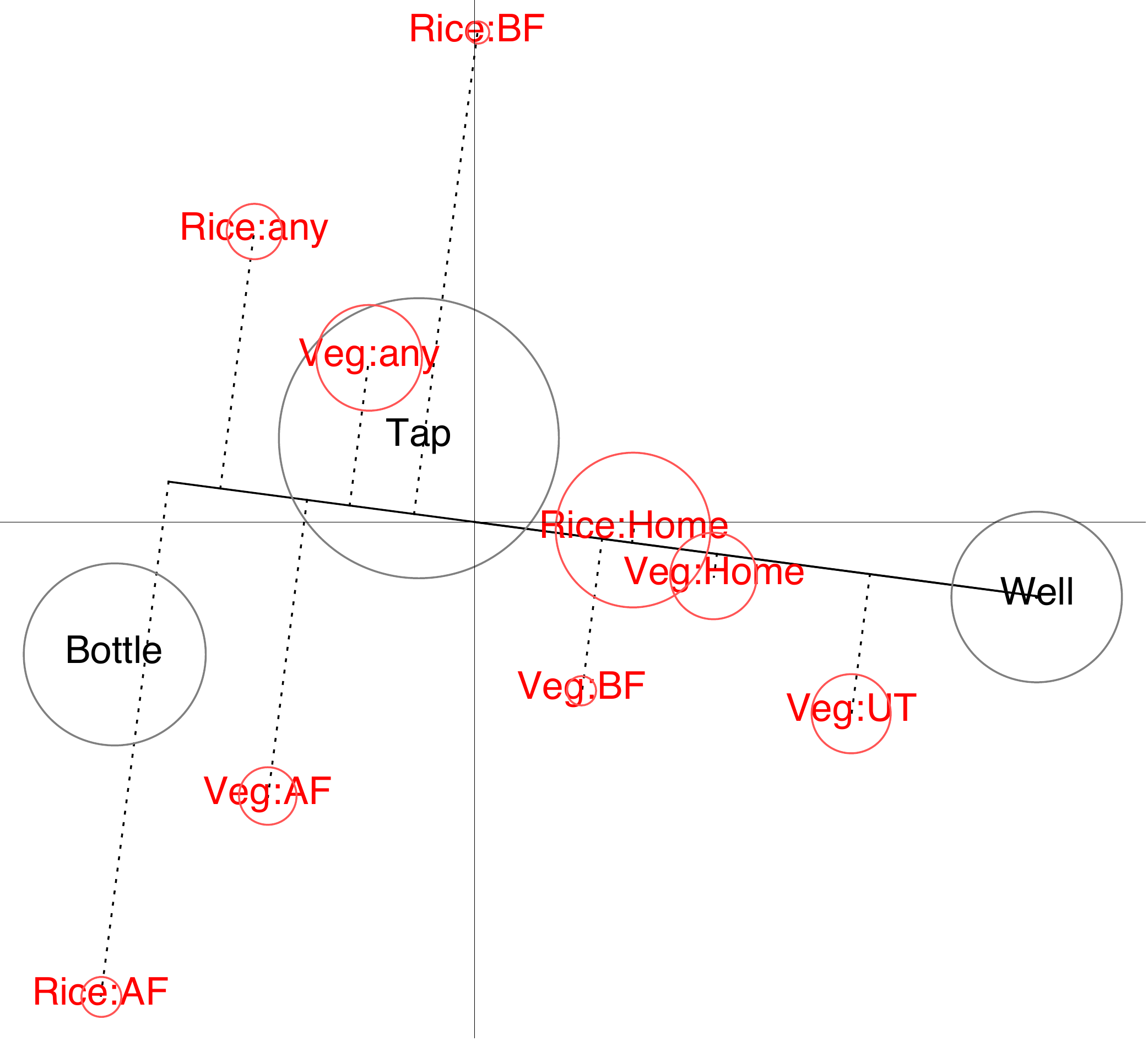}}
\caption{\label{fig:bottledwater} A graphical representation of the result of a correspondence analysis performed on the cross table (Table~\ref{tab:cross}), showing the correlation between the choice of ``bottled water'' (left) and ``well water'' (right) with the choice of rice and vegetables. For details, see text. Abbreviations are, AF - avoid Fukushima, BF - buy Fukushima, NC - do not care about the source, Home - local/home, UT - eat untested.}
\end{figure}

In order to confirm this, we performed a correspondence analysis~\cite{ca} using a standard mathematical procedure of row principal scoring~\cite{ca1}, the result of which is graphically displayed in Fig.~\ref{fig:bottledwater}.  Black circles are for the choice of drinking water and red circles are for the choice of rice and vegetables.
In this graph, the area of the red circles are proportional to the entries in the last row of Table~\ref{tab:cross}.  The solid line drawn through the ``Bottle'' circle represent the principal axes for those who buy bottled water, and the intersection of the dotted lines drawn perpendicular to the solid line indicate the relative ``distance'' of the choice of rice/vegetable to the choice of bottled water. As shown, those who buy bottled water tend to avoid Fukushima rice/vegetables. 

\section{Conclusions}
In conclusion, comprehensive whole-body counter surveys of Miharu Town school children have been conducted for four consecutive years, in 2011--2014. This represents the only long-term, sampling-bias-free internal contamination study conducted after the Fukushima Dai-ichi accident. For the first time, in 2014, the Babyscan, which has a low $^{134/137}$Cs MDA of $< 50$~Bq/body, was used to screen children shorter than 130~cm.
No child in this group was found to have detectable level of radiocesium. Using the MDAs, the upper limits of daily intake of radiocesium were estimated for each child. For those screened with the Babyscan, the upper limits were found to be $\lesssim 1$~Bq/day for $^{137}$Cs, which is consistent with the results of other studies, such as the duplicate diet study conducted by Co-op Fukushima~\cite{coop}.
Analysis of the questionnaire filled out by the parents of the children regarding their families' food and water consumption revealed that the majority of Miharu children regularly consume local or home-grown rice and vegetables. This has not resulted in an increased body burden, however, compared to children who consume only non-Fukushima food. This however does not increase the body burden. Correspondence analysis of the questionnaire results indicated that families which buy bottled water also tend to avoid Fukushima produce.  Differences in risk perception clearly exist among populations living in different parts of Fukushima. Similar correspondence analysis conducted for different Fukushima communities may provide useful information for clarifying concerns related to food and other perceived risks, and planning appropriate interventions.


\end{document}